\documentclass[12pt]{spieman}
\usepackage{amsmath,amsfonts,amssymb}
\usepackage{graphicx}
\usepackage{setspace}
\usepackage{tocloft}

\usepackage{amsmath,amssymb}
\usepackage[normalem]{ulem}
\usepackage{float}  
\usepackage{soul}
\usepackage{upgreek}
\usepackage{textcomp}

\title{Resonant microtaper leaky-mode computational spectropolarimetry with tens of femtometers spectral resolution and full stokes measurement}

\author[1,$\dagger$,*]{Yangyang Wan}
\author[1,$\dagger$]{QianYu Zhou}
\author[1]{Lin Ma}
\author[1,*]{Xinyu Fan}
\author[1]{Zuyuan He}

\affil[1]{State Key Laboratory of Photonics and Communications, Shanghai Jiao Tong University, Shanghai 200240, China.}
\affil[$\dagger$]{These authors contributed equally to this work.}

\cftpagenumbersoff{figure}
\cftpagenumbersoff{table} 
\begin{document} 
\maketitle

\begin{abstract}
% abstract, introduction最后一段以及conclusion一段，在最后正文内容完全确定后，我会再进行编写以及重新修改。
Emerging computational measurement techniques for acquiring multi-dimensional optical field information, such as spectrum and polarization, are rapidly advancing and offer promising solutions for realizing high-performance miniature systems. 
The performance of these computational measurement approaches is critically influenced by the choice of random media, yet a general framework for evaluating different implementations remains absent.
Here, we propose a universal analytical model for computational measurement systems and reveal that the system resolution is fundamentally determined by the maximum optical path difference (OPD) permitted within the random medium.
Building on this theoretical foundation, we present a resonant leaky-mode (RLM) spectropolarimeter that achieves a record high resolution-footprint-product metric. 
The RLM spectropolarimeter leverages the complex coupling between leaky modes in a tapered coreless optical fiber and whispering-gallery modes (WGM) of microsphere to significantly enhance the maximum OPD within a compact footprint. 
We simultaneously achieve an ultrahigh spectral resolution of 0.02 pm, a spectral measurement bandwidth of 150 nm, and full-Stokes polarization measurement with an accuracy of $4.732 \times 10^{-6}$, all within a sub-square-millimeter footprint. 
The proposed theoretical model clarifies the key factors governing the performance of computational measurement systems based on random media and may inspires novel design of advanced computational measurement systems for optical field. 
The demonstrated RLM spectropolarimeter offers a potential approach for highly integrated, high-performance multi-dimensional optical field measurement.

\end{abstract}

% Include a list of up to six keywords after the abstract
\keywords{computational spectropolarimetry, reconstructive spectrometer, speckle}

% Include email contact information for corresponding author
{\noindent \footnotesize\textbf{*}Yangyang Wan, \linkable{YangyangWan@sjtu.edu.cn}; Xinyu Fan, \linkable{fan.xinyu@sjtu.edu.cn} }

\begin{spacing}{1}   % use double spacing for rest of manuscript

\section{Introduction}
The multi-dimensional properties of complex optical field, including intensity, polarization and spectral, contain rich information that is fundamental to understanding light-matter interactions\cite{berry2023singularities,weiner2008light}. 
The measurement of multi-dimensional optical field information plays a vital role in modern optics and photonics research.
Polarimetric measurements, for instance, can reveal sample components \cite{he2021polarisation}, surface textures of objects\cite{tyo2006review}, and handedness of chiral molecules\cite{vinegrad2018determination}, while spectral analysis provides insights into chemical composition\cite{schliesser2005frequency}, particle velocities\cite{liu2016spectrometer}, and chemical reaction kinetics\cite{zaera2014new}.
As a more comprehensive measurement method, spectropolarimetry integrates simultaneous spectral and polarization measurements, finding extensive applications in remote sensing\cite{glenar1994acousto}, optical communications\cite{hsiang2023novel}, biochemical analysis\cite{yermolenko2009spectropolarimetry}, and astrophysics\cite{trujillo2002astrophysical}.

Conventional spectropolarimetry methods can be categorized into three types based on their principles: division-of-time, division-of-space, and channeled methods. 
In the division-of-time method, the intensities of distinct polarization and spectral components are sequentially measured using tunable filters that selectively transmit specific polarization states or wavelength components\cite{liu2021high,Meng:14,Alali:13}. 
In the division-of-space method, spectrum and polarization information are separated in the spatial domain through polarization beam splitters and dispersive elements, such as gratings and prisms, followed by parallel measurement\cite{Peinado:13,Perreault:13,delRio-Lima:24}.
In the channeled method, polarization information is modulated onto spectral or spatial carrier frequencies to enable extraction of the desired polarization states, and multiple-order retarders are utilized to generate spectral carrier frequencies coupled with spectrometer-based detection for spectral measurement\cite{Alenin:14,Vaughn:18,Chen:23,Tyo:23,Huang:24,Zhao:23,Hu:24}.
These conventional methods typically suffer from bulky size due to inherent scanning devices, multiple measurement paths, and complex integration mechanisms for combined spectral and polarization measurement, which limits their applications in miniaturized platforms\cite{chen2022advanced,dong2020polarimeters}.
In recent years, a metasurface-based method has emerged for realization of ultra-compact spectropolarimetry\cite{ni2022computational,chen2024imaging,ding2017beam,chen2016integrated}.
By engineering metasurface structure to diffract distinct polarization states and spectral components into separate spatial directions, this method enables direct far-field detection of spectropolarimetric information through single-shot measurement.
Another compact method employing a custom-designed diffractive optical element integrated with a vortex retarder transforms incident light into structured vector beams\cite{gao2025broadband}.
This method achieves simultaneous measurement of spectral and polarization information through the analysis of concentric circles with different radii and azimuthal intensity variations in the structured optical field.
However, current methods are limited in achieving high accuracy and broadband spectropolarimetric measurement with a simple and miniaturized setup.

Recent advances in computational spectrometer and polarimeter have demonstrated remarkable capabilities for high-resolution spectral and polarization measurement with compact configurations\cite{yang2021miniaturization,xue2024advances,wang2023simple,hao2024cartesian}. 
After light field propagates through natural or engineered random media, such as pearl\cite{kwak2020pearl}, multi-mode fiber\cite{redding2013all}, integrating sphere\cite{metzger2017harnessing}, and cascaded Mach-Zehnder interferometer\cite{yao2023integrated}, it generates unique speckle patterns in the time or spatial domain.
These speckle patterns inherently encode multi-dimensional information of incident light, including polarization and spectral characteristics, thereby enabling the development of diverse computational spectrometers or polarimeters based on various random media\cite{wang2023simple}.
Notably, multi-mode fiber-based high-dimensional light analyzer have achieved simultaneous wavelength determination and polarization characterization\cite{xiong2025multimode,zhou2024all}. 
Despite the demonstrated high performance of random media based computational spectrometer and polarimeter, the underlying physical mechanisms enabling high-resolution capabilities remain insufficiently explored, which impedes the structural optimization of random media for higher performance within small size.
Therefore, realizing sub-picometer spectral resolution with full stokes polarization measurement within a millimeter-scale footprint remains a significant challenge.

In this work, we demonstrate a miniaturized high-performance spectropolarimeter based on microtaper leaky-mode and whispering gallery mode (WGM) microcavity.
A theoretical model for speckle-based spectral and polarimetric measurement is established, which reveals the fundamental relationship between resolution and random media characteristics in computational spectrometer or polarimeter.
The proposed general model is verified in computational spectrometers based on different random media.
Guided by this model, a miniaturized resonant leaky-mode (RLM) spectropolarimeter is proposed by employing microtaper coreless fiber and WGM microsphere resonator to achieve enhanced resolution within confined device size.
The optimized architecture of the RLM spectropolarimeter generates intricate speckle patterns through complex interference and scattering between microtaper-induced leaky modes and microsphere resonances. 
Since the speckle pattern has a one-to-one correspondence with polarization and spectral components, a reconstruction algorithm is designed to simultaneous reconstruct spectral and polarization information from the speckle pattern with 0.02 pm spectral resolution and $4.732 \times 10^{-6}$ polarization accuracy.
The RLM spectropolarimeter exhibits a sub-square-millimeter footprint while achieving performance metrics with a bandwidth-to-resolution ratio of $7.5 \times 10^6$ and a record low resolution-footprint product of 5 nm $\times$ \textmu $\mathrm{m}^{2}$ for spectral measurement, which shows potential for broad applications in high-precision spectral polarization measurements.

\section{Results}

\subsection{Theoretical model for speckle correlation in random media}

Computational spectrometer and polarimeter acquires polarization and spectral information of the incident light field from speckle patterns generated from random media.
Various implementations of computational spectrometers and polarimeters utilizing different random media have emerged in recent years, exhibiting varying performance depending on the employed random media structures.
While previous studies have discussed relationships between structures parameters of specific random media and the reconstruction performance, an analytical model applicable to all random media remains absent.
Here, we attempt to develop a theoretical model to explore the fundamental determinants of resolution across diverse computational spectrometer and polarimeter implementations, which may provide guidance for the structural optimization of random media in computational systems.

In general implementations of computational spectrometers and polarimeters, speckle patterns arise from the interference of multiple coherent light fields generated as the light source propagates through the random medium, which may involves processes such as multiple reflection, scattering, or diverse spatial transmission modes.
Since the nonlinear effect is usually not involved in the process of speckle formation, the distinctions between multiple coherent light fields generated by any random medium can generally be attributed to variations in optical path length.
Therefore, the propagation of a light field through any random medium can be effectively represented as transmission through multiple paths with distinct lengths (See in the Supplementary Materials Fig. S1). 
For analytical purposes, the light field at position $s_0$ after passing through the random medium can be expressed as
\begin{equation}
E({s_0}) = \frac{1}{{\sqrt M }}a{E_0}\exp (2ik{Z_0})\sum\limits_m {\exp (2ik{D_m})} 
\label{eq1}
\end{equation}
where $M$ is the total number of optical paths, $a$ is the optical path attenuation coefficient, $E_0$ is the initial amplitude of the incident light, $k$ is the propagation constant, $Z_0$ is the average optical path length of multiple light fields, $D_m$ is the difference between the $m$th path length and $Z_0$.
The intensity at position $s_0$ is $I({s_0}) = {\left| {E({s_0})} \right|^2}$.
According to Eq.(1) and the reconstruction process, it can be deduced that the maximum number of measurable channels in the system is equal to the number of different optical path lengths in the random medium (Details in the Supplementary Materials S1.1).
The distinct propagation constants induced by variations in spectra or polarization states lead to the generation of dissimilar speckle patterns. 
The cross-correlation function is typically employed to evaluate the degree of similarity between speckle patterns, serving as a critical metric for determining the discernibility of corresponding spectral or polarization changes.
The correlation function between two speckle patterns with a propagation constant difference of $\Delta k$ can be expressed as (Details in the Supplementary Materials S1.1)
\begin{equation}
C(\Delta k) = \frac{{\sin {{(\Delta k\Delta {D_{\max }})}^2}}}{{{{(\Delta k\Delta {D_{\max }})}^2}}}
\label{eq2}
\end{equation}
where $D_{\max }$ is the maximum optical path difference between optical paths.
When the variation of propagation constant and the maximum optical path difference meet $\Delta k\Delta {D_{\max }} = \pi$, $C(\Delta k)$ decreases to 0 and the two speckle patterns can be completely distinguished to realize the measurement of propagation constant variation.
Therefore, the minimum variation of optical wave vector that can be measured by computational spectrometer or polarimeter is inversely proportional to the maximum optical path difference in random medium.
Taking spectral measurement as an example, through the above theoretical model, the final distinguishable optical frequency variation is $\Delta f = c/2\Delta {D_{\max }}$, where $c$ is the speed of light.
From this model, it is clear that the resolution of the computational spectrometer and polarizer ultimately depends on the maximum optical path difference allowed in the adopted random medium.
Both classic multi-mode fiber-based spectrometer relying on intermodal interference and Rayleigh speckle based spectrometer exploiting multiple scattering interference in single-mode fiber are consistent with the proposed model (Details in the Supplementary Materials S1.2).
The resolution of a Fourier-transform spectrometer (FTS) is also constrained by the maximum optical path difference, revealing that the computational spectrometer can be regards as a kind of FTS with randomized sampling intervals.

\subsection{Working principle of RLM spectropolarimeter}

\begin{figure*}[tb]
\centering
\includegraphics[width=16cm]{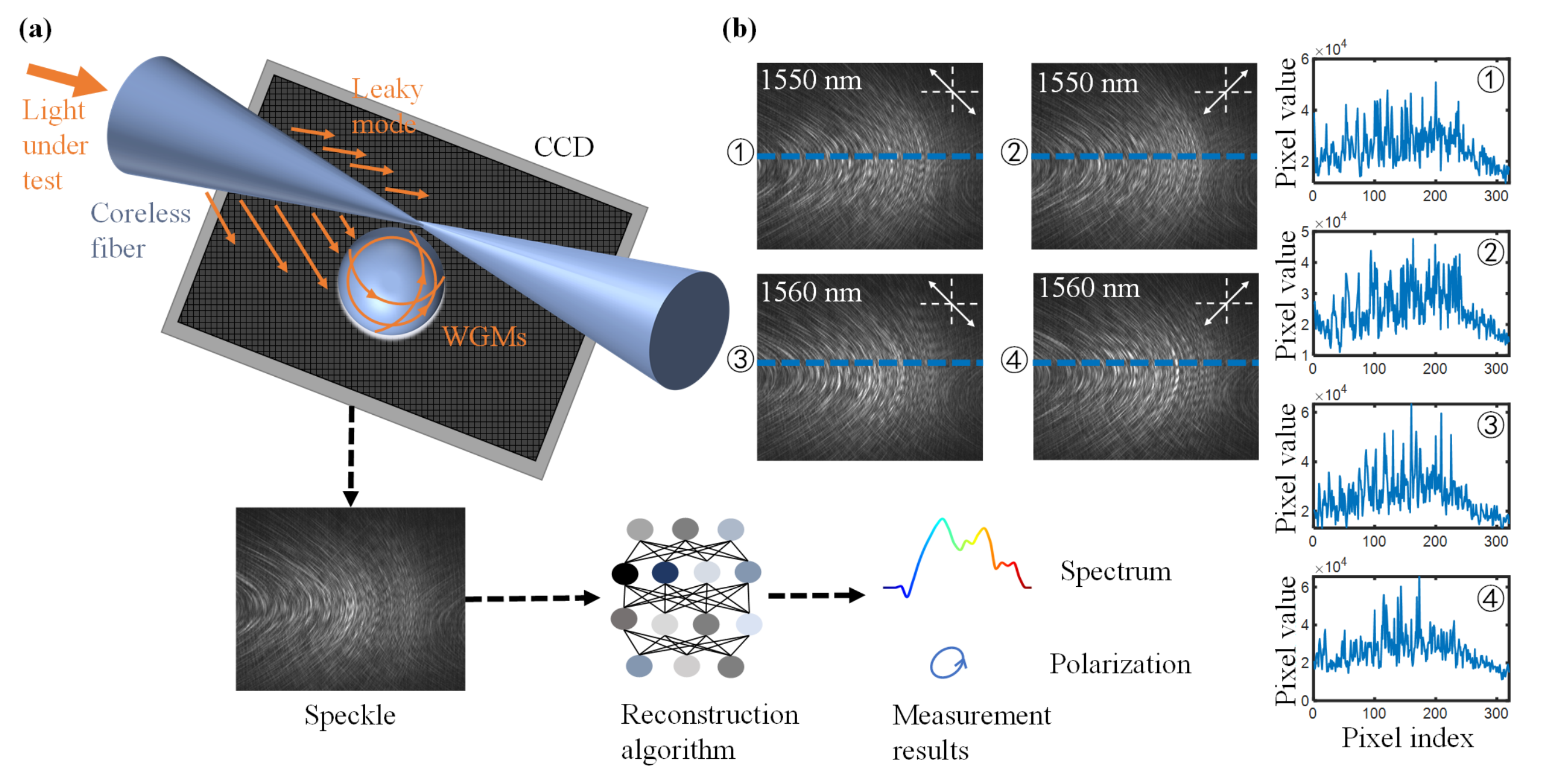}
\caption{\textbf{Principle of resonant leaky-mode (RLM) spectropolarimeter.} \textbf{(a)} Schematic diagram of RLM spectropolarimeter. The speckle patterns are generated by interference and scattering of numerous leaky modes in the coreless tapered fiber and whispering gallery mode (WGM) microsphere resonator. Spectrum and polarization information can be recovered from speckle patterns by reconstruction algorithm. \textbf{(b)} Examples of speckle patterns at different wavelengths and polarization states. The curves on the right panel of the figure represent the pixel values along the central row of each speckle pattern, highlighting the differences between the speckles.
}
\label{fig1}
\end{figure*}

Theoretical analysis indicates that enhancing the performance of computational spectropolarimetry relies on maximizing the optical path difference (OPD) within the system. 
Therefore, we propose a RLM spectropolarimeter using coreless fiber (CLF) and WGM microcavity, and the basic working principle is shown in Fig. 1(a).
The original idea is to reconstruct the spectral and polarization information of the incident light field from the speckle formed by the interference between the leaky modes of the tapered fiber. 
Under guidance of the theory, a coreless fiber is adopted to replace conventional multi-mode fiber (MMF), which is usually used for the generation of leakage mode speckle. 
By removing the coating layer and utilizing the refractive index contrast between air and coreless fiber, the incident lightwave propagates and expands freely along the CLF.
Since the optical field freely expands within the CLF, a significantly larger number of transmission modes can be supported. 
Taking the 250/500 \textmu m diameter CLF used in the experiment as an example, it supports approximately 128000 modes, while the common step-index MMF with 105 \textmu m core diameter typically supports about 1000 modes.
The propagation constant difference between the highest-order mode and the fundamental mode in the CLF is theoretically about $1.8 \times 10^6$ rad/m, which is more than one order of magnitude higher than $6.5 \times 10^4$ rad/m in step-index MMF.
Compared with step-index MMF, the greater OPD between transmission modes in a CLF of the same length makes it more suitable for miniaturized high-resolution computational spectropolarimetry. 
Additionally, since the measurement bandwidth of computational spectropolarimetry is constrained by the number of distinct optical path lengths within the system, the abundant modes in CLF can substantially extend the measurement bandwidth.

In order to further enhance the maximum OPD within the system, a WGM microsphere is introduced to resonate the leakage mode. 
The high-Q resonant modes of the WGM microsphere used in experiment with diameter of 425 \textmu m typically exhibit an angular quantum number ($l$) of about 860. 
Compared to low order leaky modes, higher order leaky modes have larger angular quantum numbers and are easier to match with the spatial distribution of high $l$ value modes in WGM microsphere.
% Additionally, the evanescent field of the tapered fiber exhibits stronger intensity for higher order modes. 
By adjusting the position of WGM microsphere, higher order leaky modes are selectively coupled into the microresonator and underwent multiple resonances, further increasing the OPD between high order and low order leaky modes.
With the incorporation of WGM microsphere, the speckle pattern generated by the interference and scattering between leaky mode fields in tapered CLF is more sensitive to the changes in spectrum and polarization of the incident light field.
As shown in Fig. 1(b), the received speckle patterns have distinct variations when either the wavelength or polarization state of the incident light is changed.
The pixel values along the central row of each speckle pattern are plotted on the right side of the figure, illustrating the differences among these various speckles.
The spectral and polarization information of interest can be computationally reconstructed from these speckle patterns using reconstruction algorithm. 
Here we adopt a convex (CVX) optimization algorithm to acquire spectrum and polarization information from speckle patterns (more details in Method).

\begin{figure*}[htbp]
\centering
\includegraphics[width=16cm]{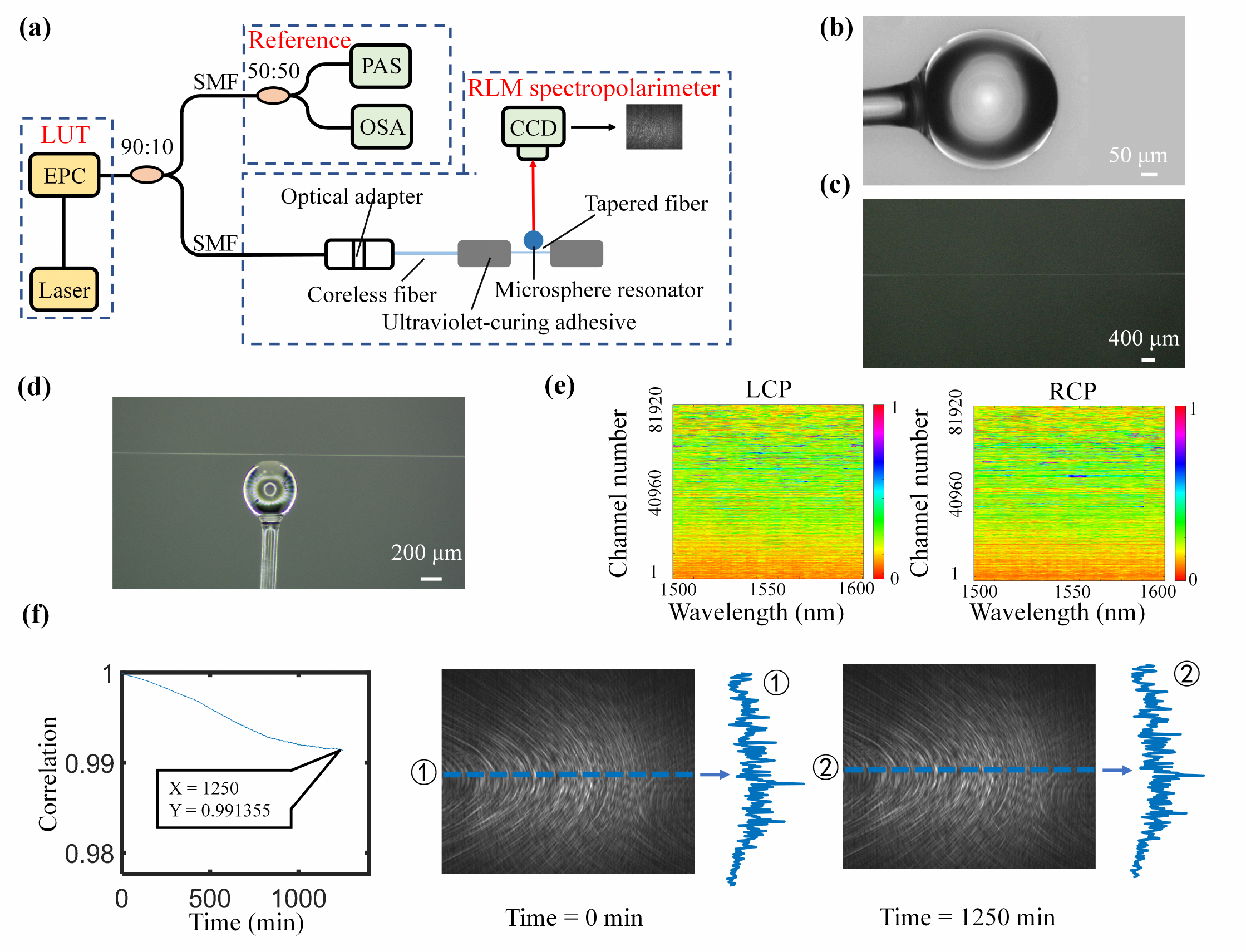}
\caption{\textbf{Experiment setup and fabricated devices.}
\textbf{(a)} The experimental setup of the proposed RLM spectropolarimeter. EPC: electric polarization controller. LUT: lightwave under test. SMF: single-mode fiber. OSA: optical spectrum analyzer. PAS: polarization analysis system. CCD: charge-coupled device.
\textbf{(b)} Microscope image of the fabricated WGM microsphere resonator with a diameter of 425 \textmu m. It is homemade with a standard commercial 125 \textmu m single-mode fiber.
\textbf{(c)} Microscope image of the tapered coreless fiber with a taper region length of 15 mm and a waist region diameter of 1 \textmu m. It is drawn under non-adiabatic condition with a 250/500 \textmu m coreless fiber. The non-adiabatic condition ensures the taper region to maximize the leakage of the incident lightwave field (More details can be found in Supplementary Material). 
\textbf{(d)} Microscope images of the structure of the proposed RLM spectropolarimeter.
\textbf{(e)} Measured transmission matrix of the proposed spectropolarimeter under different polarization states. The right section shows the matrix under the right-handed circular polarization state (RCP), while the left section corresponds to left-handed circular polarization state (LCP). The horizontal axis denotes spectral channels, and the vertical axis represents spatial channels.
\textbf{(f)} Long-term stability of the speckle pattern generated by the proposed spectropolarimeter.
The spectrum and polarization state of the incident lightwave are held constant, and the stability of the system is evaluated based on the temporal distortion of the generated speckle pattern. The temporal distortion is quantified by the cross-correlation coefficient between the speckle pattern at a given time and the initial one. A higher correlation coefficient indicates greater long-term stability. 
The speckle patterns acquired at the 0th and 1250th minutes are shown on the right side of the correlation coefficient curve, and the pixel values along the central row of each speckle pattern are displayed alongside the images.
For this system, the speckle patterns maintain a high correlation with the initial speckle after 1250 minutes (More details can be found in Supplementary Material).
}
\label{fig2}
\end{figure*}

\subsection{Fabricated devices and experimental setup}

The experimental setup of RLM spectropolarimeter is shown in Fig. 2(a).
The light under test (LUT) is divided into two paths through a 90:10 coupler. 
90$\%$ of the light enters the RLM spectropolarimeter for measurement, while the remaining enters the standard commercial polarization analysis system (PAS, General Photonics POD-201) and optical spectrum analyzer (OSA, Yokogawa AQ6370C) for reference.
In RLM spectropolarimeter, LUT generates a significant number of leaky mode optical fields after passing through the fabricated coreless tapered fiber. 
A portion of the leaky modes further couples into the WGM microsphere, forming complex resonances. 
A charge coupled device (CCD) captures the speckle pattern formed by interference among these leaky modes.
It should be noted that nearly all computational measurement systems require a calibration process prior to measurement to obtain the transmission matrix of the random medium or other prior knowledge needed for reconstruction.
During calibration, we use a frequency-tunable laser combined with an electrically controlled polarization controller (EPC) to generate a reference light source with arbitrary polarization and wavelength. 
By varying the state of reference light, we acquire corresponding speckle patterns at different wavelengths and polarization states to establish the prior knowledge required for the reconstruction algorithm.
Subsequently, during measurement, spectral and polarization information can be recovered by the reconstruction algorithm from the measured speckle patterns.

The WGM microsphere resonator used in the experiment is fabricated using the fiber fusion method. 
By applying arc discharge from a fiber splicer to thermally process the fiber tip, the molten quartz material forms a spherical structure under surface tension. 
Fig. 2(b) shows a microsphere with diameter of 425 \textmu m fabricated from a standard 125 \textmu m single-mode fiber.
WGM microsphere generated by the fiber fusion method often exhibits non-ideal sphericity. 
Such asymmetry breaks degeneracy, causing broadening and splitting of whispering-gallery resonance peaks. 
While this imperfection is typically undesirable in most WGM based sensing systems, it benefits the RLM by increasing both the phase-matching points between higher order leaky mode optical fields and WGMs and the complexity of leaky mode resonances, while simultaneously reducing the efficiency of low order leaky mode coupling to maximize the OPD between low order and high order leaky modes.

The tapered coreless fiber is fabricated via hydrogen-oxygen flame heating and pulling, yielding a waist region approximately 1 \textmu m in diameter as shown in Fig. 2(c) (Details in Supplementary Material).
A smaller waist diameter increases the number of leaky modes, thereby enhancing both the resolution and measurement bandwidth of the RLM spectropolarimeter.
After precisely adjusting the position between the coreless tapered fiber and the WGM microsphere using a precision displacement stage as shown in Fig. 2(d), the leaky modes can coupled into the WGM microsphere.
Speckle patterns acquired under right-handed circular polarization (RCP) and left-handed circular polarization (LCP) across the 1500 nm to 1600 nm wavelength range are shown in Fig. 2(e), serving as a representative template.
Distinct pattern features corresponding to varying wavelengths and polarization states demonstrate the feasibility of reconstructing spectral and polarization information from the speckle.

High-precision measurement instruments are typically sensitive to environmental perturbation. 
In our system, the footprint of the random medium is approximately 500 \textmu m $\times$ 500 \textmu m. 
This compact size endows the system with good long-term stability and environmental perturbation immunity.
As shown in Fig. 2(f), we evaluate the long-term stability of system using a frequency-stabilized laser (daily drift $\textless$ 100 kHz) as the light source.
Continuous speckle pattern acquisition from this fixed-wavelength source enabled environmental interference assessment through cross-correlation analysis with the initial speckle pattern. 
The result shows that the system maintains a speckle correlation coefficient above 0.99 for over 1250 minutes in laboratory environment without observable pattern distortion.
The speckle patterns received at the beginning of the experiment and obtained after 1250 minutes are shown on the right side of the correlation coefficient curve.
The corresponding pixel values along the central row of each speckle pattern are also displayed alongside the images for comparison.
More long-term stability demonstrations can be found in Supplementary Material.

\begin{figure*}[htbp]
\centering
\includegraphics[width=16cm]{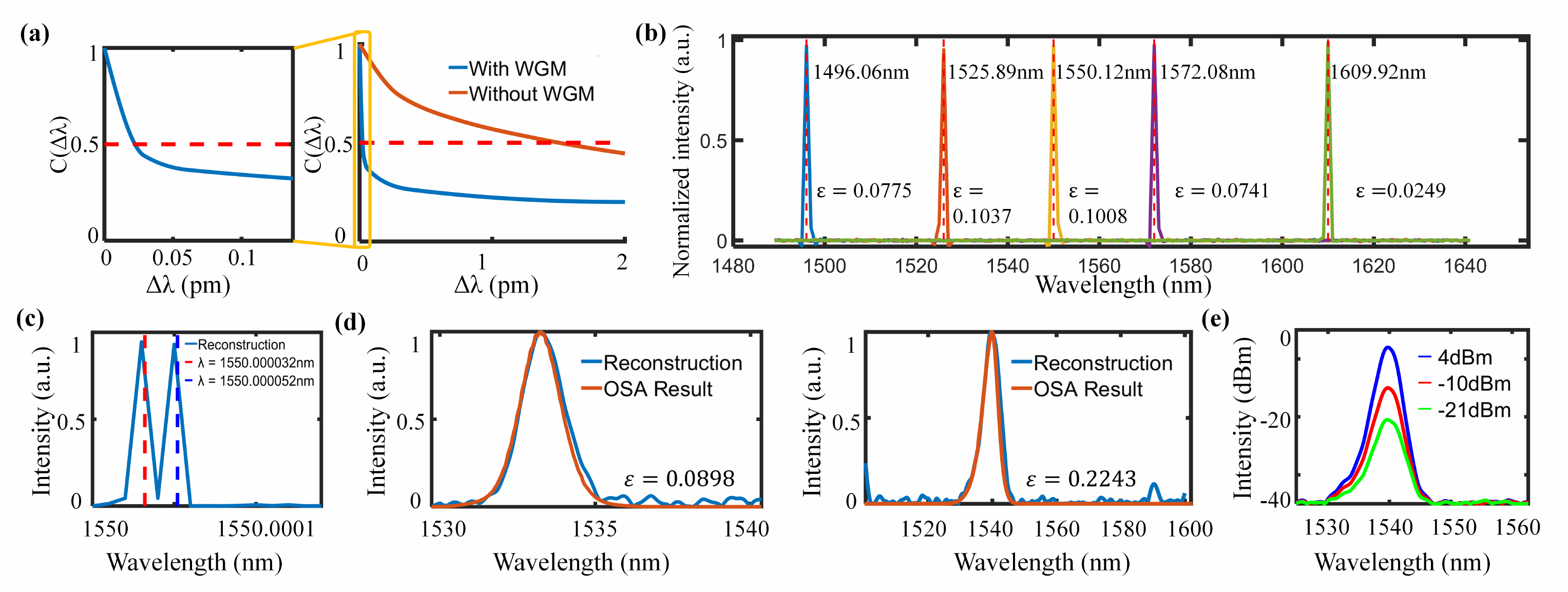}
\caption{\textbf{Spectra measurement results.} 
\textbf{(a)} Wavelength-dependent correlation curves with and without the WGM microsphere resonator. The wavelength shift corresponding to a correlation coefficient of 0.5 is defined as the theoretical spectral resolution of the system. With the incorporation of the WGM microsphere resonator, the spectral resolution improves significantly from 1.5 pm to 0.02 pm. The enlarged image on the left panel shows more details of the correlation curve with the WGM microsphere resonator.
\textbf{(b)} Reconstructed narrowband spectrum with known peak positions at 1496.06 nm, 1525.89 nm, 1550.12 nm, 1572.08 nm, and 1609.92 nm. 
Solid and dashed lines represent the reconstructed spectrum and the ground truth, respectively. The relative error coefficients $\varepsilon$ for each peak are annotated next to the corresponding reconstructed spectrum. The relative error $\varepsilon$ is defined as $\varepsilon=\|S-S_r\|_2 /\|S\|_2$, where $S$ denotes the actual spectrum and $S_r$ is the reconstructed spectrum.
\textbf{(c)} Reconstructed spectrum of dual narrowband peaks separated by 0.02 pm.
The red and blue dashed line represent two incident narrowband lightwave with center wavelengths of 1550.000032 nm and 1550.000052 nm, respectively. The solid line corresponds to the reconstructed spectrum, which distinguishs these two spectral components.
\textbf{(d)} Reconstructed spectrum and corresponding relative error for different input broadband lightwave sources.
The blue solid line represents the reconstructed spectrum, while the red solid line denotes the reference spectrum obtained by a commercial spectrometer, which is considered the ground truth.
The left section shows a broadband lightwave with a full width at half maxima (FWHM) of approximately 3 nm over a 10 nm span, and the right section shows a broadband lightwave with a FWHM approximately 12 nm within 100 nm.
The amplified spontaneous emission (ASE) noise generated by an erbium-doped fiber amplifier (EDFA) is shaped by a waveshaper to form the broadband input, which then enters into the system for measurement.
\textbf{(e)} Broadband spectrum reconstruction under varying input optical powers ranging from -4 dBm to -21 dBm, with a center wavelength of 1539.8 nm.
Further discussion about the reconstruction performance for both narrowband and broadband lightwave fields at low input optical power is provided in Supplementary Material.
}

\label{fig3}
\end{figure*}

\subsection{Experimental measurement results}

\begin{figure*}[htbp]
\centering
\includegraphics[width=16cm]{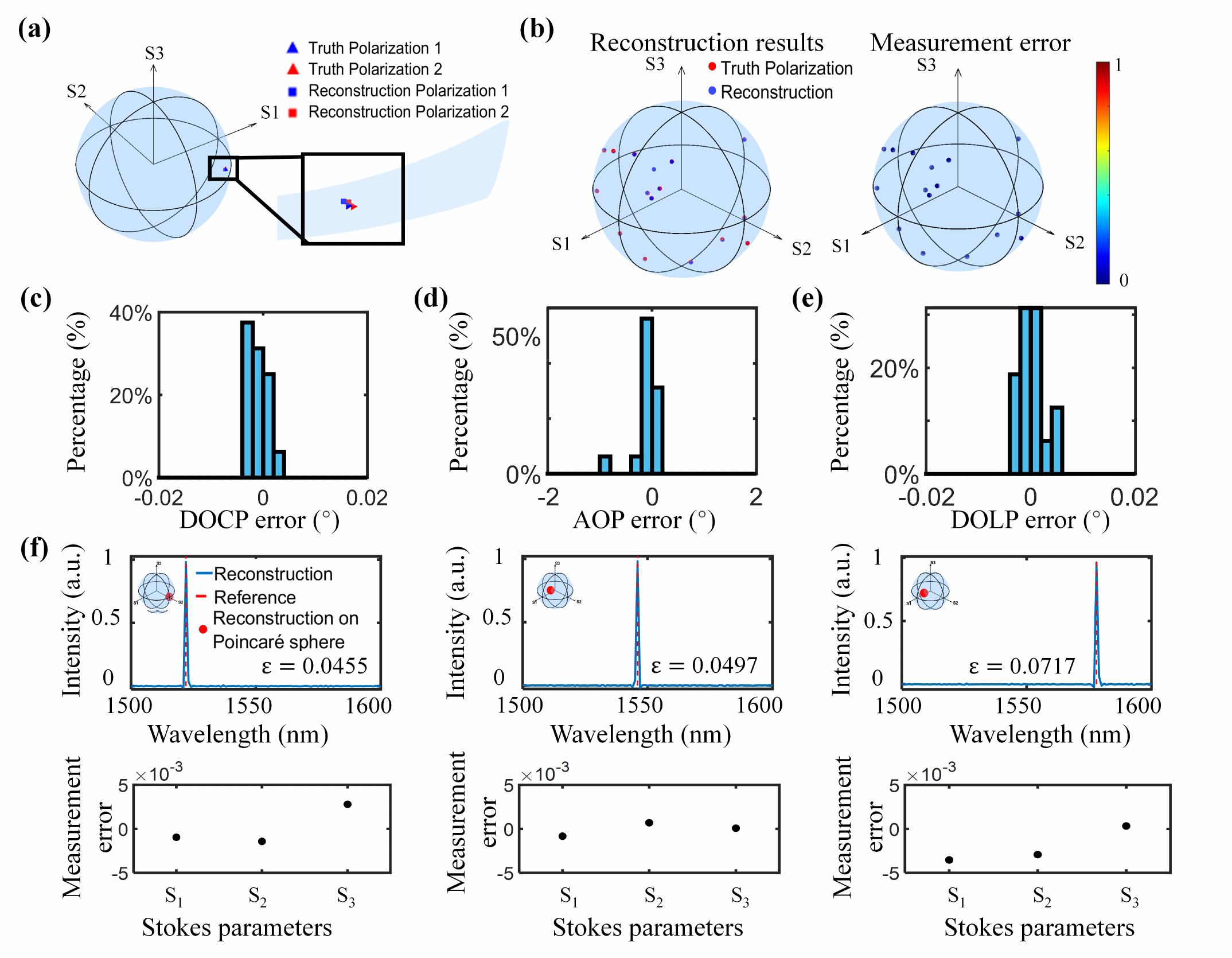}
\caption{\textbf{Polarization measurement results and spectropolarimetric measurements.} 
\textbf{(a)} Polarization resolution measurement results. Two polarization states, with Stokes vectors [0.582, -0.796, 0.167] and [0.587, -0.791, 0.173], are accurately reconstructed and clearly resolved. The blue and red triangles donate the ground-truth values, while the blue and red squares represent the corresponding reconstructed results. The relative distance and orientation between the true and reconstructed polarization states remain consistent, demonstrating that the system is capable of distinguishing polarization states with such differences. For this system, the polarization resolution is quantified as 0.00149, which corresponds to the spherical distance between the two polarization states on the Poincaré sphere.
\textbf{(b)} The left section shows the reconstructed polarization states (red dots) and the corresponding ground truth values (blue dots) on the Poincaré sphere. The right section donates the reconstruction error corresponding to each polarization state shown on the left. The color indicates the value of the polarization measurement error, which is quantified by the $L_2$-norm of the deviation between the reconstructed results and the ground truth.
\textbf{(c)} Distributions of measurement errors for the angle of polarization (AOP), degree of linear polarization (DOLP), and degree of circular polarization (DOCP) corresponding to the reconstructed results in (b). X-axes indicate the error values, and y-axes donate the corresponding percentage.
\textbf{(d)} The upper section shows the results of the simultaneous measurement for spectrum and polarization state. The narrowband incident lightwave hold center wavelengths of 1522.03 nm, 1545.96 nm, and 1578.08 nm, respectively. The reconstructed polarization states are shown on the Poincaré sphere in the upper-left corner. The lower section presents the corresponding reconstruction errors of the Stokes vectors.
}
\label{fig4}
\end{figure*}

The spectral measurement results of the proposed RLM spectropolarimeter are shown in Fig. 3. 
Typically, speckle-based spectrometers employ cross-correlation coefficient curves to quantify speckle variation with wavelength changes, where the wavelength shift corresponding to a correlation coefficient drop to 0.5 defines the resolution. 
Although advanced reconstruction algorithms such as machine learning methods can further improve resolution, the correlation-based resolution definition provides a universally valid metric for resolution characterization.
Fig. 3(a) compares wavelength-dependent correlation curves with and without the WGM microsphere structure. 
When WGM-enhanced resonance of leaky modes is implemented, the resolution improves from 1.5 pm to 0.02 pm. 
Fig. 3(b) demonstrates wavelength measurement results across different spectral bands.
Narrowband peak with center wavelengths at 1496.06 nm, 1525.89 nm, 1550.12 nm, 1572.08 nm, and 1609.92 nm are reconstructed successfully with low relative reconstruction error.
% Solid and dashed lines are the respective reconstruction results and the reference spectrum.
% The relative error coefficients $\varepsilon$ are positioned next to the corresponding curves.
The relative reconstruction error $\varepsilon$ is obtained by calculating the mean square error between the reconstructed spectrum and the true spectrum, which can be expressed as:
% $\varepsilon=\frac{\left\|S-S_r\right\|_2}{\|S\|_2}$
$\varepsilon={\left\|S-S_r\right\|_2} / {\|S\|_2}$
where $S$ denotes the spectrum obtained from OSA, and $S_r$ is the reconstructed spectrum.

The spectral resolution is determined by successfully resolving two spectral lines separated by 0.02 pm, as shown in Fig. 3(c). 
Moreover, Fig. 3(d) shows the measurement results of different broadband light sources and corresponding reconstruction errors, when compared with commercial spectrometer readings as ground truth, are 0.0898 and 0.2243, respectively.
% The blue line is the reconstructed result, and the red line denotes the ground truth.
% The figure on the left shows the reconstruction of a broadband lightwave source with a full width at half maxima (FWHM) of approximately 3 nm over a 10 nm spectral span. 
% Meanwhile, the figure on the right illustrates the reconstruction of a broadband source with an FWHM of 12 nm across a 100 nm range.
Broad spectrum sources with bandwidths of 3 nm and 12 nm can be accurately reconstructed.
Notably, the WGM microsphere enhances evanescent field intensity, improving power sensitivity of the proposed system (Details in the Supplementary Materials). 
Measurements of broadband light at varying optical power levels are presented in Fig. 3(e). 
The system accurately reconstructs spectrum at peak powers of -21 dBm, demonstrating a measurement sensitivity below -30 dBm.

To evaluate the capability of the RLM spectropolarimeter to distinguish different polarization states, here we define polarization resolution as the minimum spherical distance between two resolvable polarization states on a Poincaré sphere.
Two closely spaced polarization states with Stokes vectors [0.582, -0.796, 0.167] and [0.587, -0.791, 0.173] are reconstructed separately, with results shown in Fig. 4(a).
Using the spherical distance formula $S = R \cdot \operatorname{arc} \cos [\cos \beta_1 \cos \beta_2 \cos (\alpha_1-\alpha_2)+\sin \beta_1 \sin \beta_2]$, the polarization resolution of the RLM spectropolarimeter is calculated to be 0.00149, corresponding to a polarization deviation angle of 0.538°.
Fig. 4(b) demonstrates the measurement results and errors of the polarization states across the entire Poincaré sphere, confirming the full characterization of the Stokes vector.
The polarization reconstruction error is quantified by the $L_2$-norm of the deviation between the reconstructed and reference Stokes vectors.
The maximum measurement error on the entire Poincare sphere is only $1.552 \times 10^{-5}$.
Histograms of the errors in the degree of circular polarization($\mathrm{DOCP}=\mathrm{S}_3 / \mathrm{S}_0$) , angle of polarization ($\mathrm{AOP}= 1/2 \arctan ({\mathrm{~S}_2}/{\mathrm{~S}_1})$), and degree of linear polarization ($\text { DOLP }=\sqrt{\mathrm{S}_1^2+\mathrm{S}_2^2} / \mathrm{S}_0$) between the measured and actual values, as obtained from the reconstruction process illustrated in Fig. 4(b), are shown in Fig. 4(c) - (e).
AOP describes the angle between the vibration direction of the incident lightwave field and the reference direction, DOLP presents the ratio of the linearly polarized component of the incident lightwave to the total lightwave intensity, while DOCP is the ratio of the circularly polarized component to the total intensity. The measurement errors associated with these three parameters further indicate the polarization measurement accuracy of the proposed system.
The reconstruction errors for three Stokes parameters are 0.58\%, 1.95\% and 0.30\% respectively, with an overall MSE of $4.732 \times 10^{-6}$.
The simultaneous spectral and polarization state measurement capability of RLM spectropolarimeter is also verified as shown in Fig. 4(f), showing agreement with the ground truth.
RLM can achieve accurate reconstruction under different spectral and polarization state measurements with low measurement error.

% \subsection{Direct optical frequency comb measurement results}

\section{Discussion}
A performance comparison of the state-of-the-art polarimeters and spectropolarimeters in polarization measurement is presented in Table 1.
The table demonstrates that this RLM spectropolarimeter achieves high-accuracy polarization reconstruction, and the polarization reconstruction error is nearly an order of magnitude better than the current state-of-the-art technology.

\begin{table}[ht]
% \centering  % 用这个代替 \begin{center}
\caption{Performance comparison of polarization measurement} 
\label{tab:Multimedia-Specifications}
\begin{center}      
\scalebox{0.85}{
\begin{tabular}{cccccc}
\hline
Structure   &  \begin{tabular}[c]{@{}l@{}}Polarization \\ error (MSE)\end{tabular} &       \begin{tabular}[c]{@{}l@{}}Polarization \\ deviation angle\end{tabular}  &              \begin{tabular}[c]{@{}l@{}}Error of Stokes \\  parameters\end{tabular}              & \begin{tabular}[c]{@{}l@{}}Full-Stokes\\ mansurement \end{tabular}   & Reference \\
\hline
Liquid crystal &  / & $1.5^\circ$ -- $4^\circ$ &  /  &  \checkmark & \cite{ni2022computational} \\
Liquid crystal &  $5 \times 10^{-3}$ & $7^\circ$ & $\sim$5\%  & \checkmark & \cite{zhu2023harnessing} \\
Metasurface &  $> 1 \times 10^{-2}$ & / & $>$10\%  & $\times$ &  \cite{ding2017beam} \\
Metasurface &  / & / & 7.85\%--13\%  &\checkmark&  \cite{chen2024imaging} \\
Metasurface &  / & / & /  &$\times$&   \cite{chen2016integrated} \\
Moiré quantum &  $2 \times 10^{-3}$ & $3^\circ$ & $\sim$10\%  &\checkmark   &  \cite{ma2022intelligent} \\
Moiré photonic crystal  &  0.0355 & / & /  &\checkmark&    \cite{tang2025adaptive} \\
Multi-mode fiber &  $5.3 \times 10^{-5}$ & $\boldsymbol{0.51^\circ}$ & 3.37\% (S$_1$), 1.01\% (S$_2$), 0.84\% (S$_3$)  & \checkmark &  \cite{zhou2024all} \\
Multi-mode fiber &  $1.06 \times 10^{-4}$ & / &  /  &\checkmark &  \cite{xiong2025multimode} \\
% Multi-mode fiber &  0.0258  $1.06 \times 10^{-4}$ & / &  /  &\checkmark &  \cite{xiong2025multimode} \\
%Coreless fiber  & 0.001 nm & 100 nm & $6.9 \times 10^{-6}$ & $1.1^\circ$ &\checkmark& \textbf{CLEO2024} \\
Diffractive element  &  $<10^{-4}$ & / & /   &\checkmark  &  \cite{gao2025broadband} \\
Optical crystals  &  / & / & /  &\checkmark&  \cite{wang2024exploiting} \\
Metal-dielectric hybrid&  / & / & 1.90\% (S$_1$), 2.70\% (S$_2$), 7.20\% (S$_3$)  &\checkmark&   \cite{basiri2019nature} \\
Spin–orbit interaction&  / & / & 7.40\% (S$_1$), 15.60\% (S$_2$), 11.40\% (S$_3$)  &\checkmark&   \cite{espinosa2017chip} \\
In-line metasurface &  / & / & 6.0\% (S$_1$), 5.8\% (S$_2$), 4.7\% (S$_3$)  &\checkmark&   \cite{balthasar2016ultracompact} \\
Dielectric metasurface&  / & / & 7.5\%--15\%  &\checkmark&   \cite{arbabi2018full} \\
Graphene anisotropic&  / & / & $>$3.9\%(S$_1$), $>$6.5\% (S$_2$), $>$2.5\% (S$_3$)  &\checkmark&   \cite{jung2018polarimetry} \\
Plasmonic metasurface&  / & / & 3.5\% (S$_1$), 2.5\% (S$_2$), 10.4\% (S$_3$)  &\checkmark& \cite{bai2019chip} \\
Single photon &  / & / & $\sim$1.7\%  &\checkmark& \cite{hu2022full} \\
Coreless fiber \& WGM & $\boldsymbol{4.73 \times 10^{-6}}$ & $0.53^\circ$ & \textbf{0.58\% (S$_1$), 1.95\% (S$_2$), 0.30\% (S$_3$)}  & \checkmark & \textbf{This work}\\
\hline
\end{tabular}}
\end{center}
\end{table}

\begin{figure*}[tb]
\centering
\includegraphics[width=16cm]{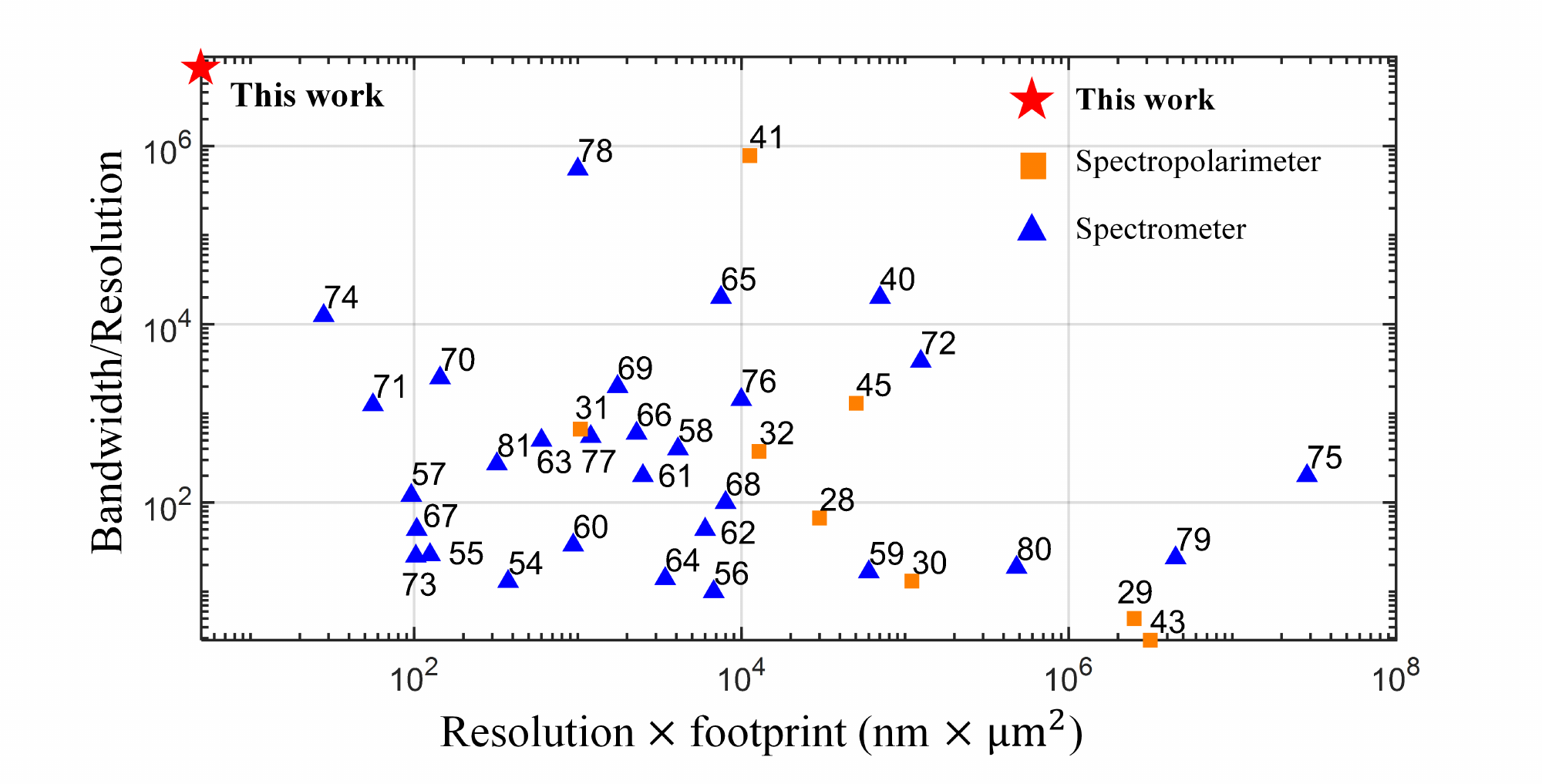}
\caption{\textbf{Performance comparison of spectrometer and spectropolarimeter in spectral measurement.} The RLM spectropolarimeter achieves a bandwidth/resolution ratio of $7.5 \times 10^6$ and a resolution $\times$ footprint of 5 nm $\times$ \textmu $\mathrm{m}^{2}$. 
}
\label{fig5}
\end{figure*}

\nocite{yang2019single}
\nocite{zheng2020chip}
\nocite{liu2020designing}
\nocite{hadibrata2021compact}
\nocite{li2021chip}
\nocite{yuan2021wavelength}
\nocite{redding2013compact}
\nocite{redding2016evanescently}
\nocite{hartmann2020waveguide}
\nocite{hartmann2020broadband}
\nocite{cheng2021generalized}
\nocite{xu2023cavity}
\nocite{zhang2021compact}
\nocite{zhang2022cascaded}
\nocite{sun2023scalable}
\nocite{zhang2022ultrahigh}
\nocite{xu2023breaking}
\nocite{xu2023integrated}
\nocite{yao2023broadband}
\nocite{zhao2024miniaturized}
\nocite{zhang2025miniaturized}
\nocite{cheng2019broadband}
\nocite{zhang2025scalable}
\nocite{uddin2024broadband}
\nocite{cen2023microtaper}
\nocite{le2007wavelength}
\nocite{velasco2013high}
\nocite{finco2024monolithic}

Fig. 5 shows a comprehensive performance comparison of leading computational spectrometers, Fourier transform spectrometers and spectropolarimeters in spectral measurements.
The spectral performance comparison evaluates two key metrics: bandwidth/resolution and resolution-footprint-product (RFP). 
The bandwidth/resolution ratio, representing the spectral dynamic range of system, indicates its capability for broadband measurement with high resolution. 
Since most spectral measurement systems exhibit a trade-off where larger dimensions enable higher spectral resolution, RFP metric is therefore introduced to evaluate achievable resolution per unit footprint. 
Benefiting from both the large propagation constant difference between high order and low order leaky modes and the complex resonance effects of WGM microcavity, RLM spectropolarimeter achieves enhanced optical path differences within a compact footprint.
With a 0.02 pm spectral resolution across 150 nm measurement bandwidth, the proposed system shows a a leading performance with a high bandwidth/resolution of $7.5 \times 10^6$ and an ultra-low RFP of 5 nm $\times$ \textmu $\mathrm{m}^{2}$.
These results demonstrate superior performance of RLM spectropolarimeter in overcoming the traditional bandwidth-resolution-size trade-off relationship in spectrum measurement, while maintaining high-performance polarization measurement capabilities.
% $\times$ \textmu $\mathrm{m}^{2}$

%讨论系统结构的影响：微球结构尺寸、微球与锥形光纤间gap的影响、CCD高度的影响。
The performance of the RLM spectropolarimeter is intrinsically linked to the WGM microsphere structure. 
Larger microsphere diameters support increased numbers of transmission modes, thereby enhancing higher-order leaky mode coupling. 
For further investigation, we fabricated microspheres with varying diameters and compared their achieved spectral resolutions (details in the Supplementary Materials).
Larger WGM microsphere resonators can support greater optical path differences, resulting in complex speckles that are more sensitive to wavelength changes and thus enabling higher spectral resolution.
The distance between the complex coupling region (composed of tapered coreless fibers and the WGM microsphere resonator) and the CCD camera also has a significant impact on system performance. 
The position of the CCD camera is adjusted using the coupled displacement platform to evaluate the spectral resolution of the proposed system at varying distances. 
The results indicate that increased distance leads to more severe degradation in spectral resolution. 
This is attributed to the reduced effective information in the speckle patterns at greater distances and complicates the reconstruction process.
The relative position between the tapered coreless fibers and WGM microsphere resonator also affects the measurement performance greatly.
Experimental results indicate that increasing separation leads to degraded spectral resolution. 
This degradation is caused by reduced coupling efficiency between the evanescent field leaked from the fiber and the WGM at larger distances, which decreases optical path differences and results in more severe spectral resolution loss.
Reasonable design and optimization of system structure may achieve higher performance in future.

The main sources of noise in the system are environmental perturbations and electrical noise.
Environmental perturbations originate from external vibrations and temperature fluctuations, while electrical noise originates from the CCD camera and its inherent thermal effects. 
All of these factors collectively contribute to errors in the reconstruction process.
Generally speaking, temperature variations are the primary factor affecting system performance, which remains a major challenge for all existing computational measurement techniques. 
Common solutions include employing temperature control technologies and implementing miniaturized systems to minimize the impact of temperature fluctuations. 
In addition, when weak optical signals are measured, the intrinsic noise of the detector can significantly affect the reconstruction results. 
This issue could be mitigated in the future through advanced noise reduction algorithms.

%总结工作

In conclusion, we present a resonant leaky-mode (RLM) computational spectropolarimeter based on tapered coreless fiber coupled with a WGM microsphere. 
By developing a universal model for computational measurement systems, we establish a generalized principle applicable to arbitrary random media: increasing the maximum optical path difference within the random medium enhances reconstruction resolution.
Guided by this theory, we employ a WGM microsphere to resonantly couple high order leaky modes in tapered coreless fiber, thereby enhancing the maximum optical path difference within a compact footprint.
The introduction of WGM resonance enhances the spectral resolution of system by 75 times to 0.02 pm, while simultaneously achieving a polarization resolution of 0.00149.
Moreover, the incorporation of WGM microsphere substantially enhances the intensity sensitivity of system.
The RLM spectropolarimeter achieves a record performance with a bandwidth/resolution ratio of $7.5 \times 10^6$ and resolution-footprint product of 5 nm $\times$ \textmu $\mathrm{m}^{2}$.
The proposed theory provides basic design principles for the implementation of computational measurement system with high resolution and broad measurement bandwidth, while the demonstrated RLM spectropolarimeter reveals potential for miniaturized, high-resolution multi-dimensional optical field measurement, offering a solution for in-situ applications. 

% $\times$ \textmu $\mathrm{m}^{2}$

\section{Methods}

\textbf{Device fabrication}

The coreless tapered fibers are fabricated using a fiber drawing machine based on hydrogen-oxygen flame fusing and mechanical pulling (details in Supplementary Materials). 
The hydrogen–oxygen generator produces hydrogen and oxygen by electrolyzing an aqueous sodium hydroxide solution. The hydrogen–oxygen flame head scans back and forth along the longitudinal axis of the optical fiber, while two stepping platforms simultaneously move outward to stretch and thin the fiber. 
% The drawing machine used in this experiment has a maximum flame head scanning range of 20 mm and a maximum stretching length of 80 mm. 
Key controllable parameters include the $\mathrm{H_2/O_2}$ flow rate, scanning width and speed of the flame head, stretching length, and stretching speed.
The tapered optical fibers used in the experiment are 250/500 $\mathrm{\upmu m}$ coreless optical fibers. Prior to drawing, the coating of the fiber is removed to prevent leakage of the incident light field through the coating layer. 
The coreless tapered fibers with a waist diameter less than 1 $\mathrm{\upmu m}$ can be fabricated.
% After cleaning with alcohol, the fiber is positioned at the pre-stretching location and covered with a dust shield to prevent contamination of the taper region caused by dust in air. The two ends of the fiber are placed on stepper motors and mechanically fixed. Once the flame head reaches the pre-stretching position, it preheats and softens the fiber. Subsequently, the flame head begins scanning according to the preset width, while the stepping motors move outward to stretch the fiber. Stretching stops once the desired length is achieved, and the diameter of the waist is less than 1 $\mathrm{\upmu m}$. 

% After tapering, a home-made mold is placed beneath the tapered coreless fiber, and both ends are secured using ultraviolet-curing adhesive, leaving the central section suspended. The mold is fabricated using 3D printing and the material is resin. To prevent the taper region from breaking due to gravity, the suspended section should be sufficiently short.

The WGM microsphere resonator employed in the experiment is manufactured using an optical fiber fusion splicer (ARC Master, FSM-100P+), and the optical fibers used are standard commercial single-mode fibers. The fusion splicer utilizes tip discharge to melt the fiber ends and form a microsphere resonator. Key adjustable parameters include discharge time, discharge intensity, motor step distance, step speed, and step acceleration.
Prior to fabrication, the coating layer of the fiber is removed. Once discharge is initiated, the motor beneath the fixture continuously advances the fiber toward the discharge region. The molten fiber gradually forms a microsphere during this process. The fabrication process ends once the preset discharge time or step distance is achieved.
Each fabricated microsphere remains attached to an unmelted section of the fiber, which facilitates coupling operations in the following. In the experiment, the coupling efficiency can be optimized by adjusting the position and orientation of the fiber connected to the microsphere resonator.

\textbf{Experiments}

A tunable source laser (TSL 8164B, Keysight), an optical spectrum analyzer (AQ6370C, Yokogawa), and a polarization analysis system (POD-201, General Photonics) are employed in the experiment for data acquisition and reference measurements. A high-speed electro-polarization controller (EPC, Sc-Lightsource) is used to rapidly switch and stabilize polarization states and communicates via the RS-232 interface standard. A charge-coupled device (CCD, Bobcat 320, Xenics) is utilized to capture speckle patterns. To ensure a consistent response of the CCD camera, it need to be calibrated prior to measurements. The tapered coreless fiber, WGM microsphere, and CCD camera are mounted on a home-made coupled displacement platform. The calibration process takes approximately 12 minutes.

For narrowband incident lightwave reconstruction, the TSL 8164B laser is used to generate lightwaves with center wavelengths of 1496.06 nm, 1525.89 nm, 1550.12 nm, 1572.08 nm, and 1609.92 nm. To demonstrate spectropolarimetric capability, lightwave at 1522.03 nm, 1545.96 nm, and 1578.08 nm with corresponding polarization states are introduced into the system for reconstruction. An ultra-stable laser is employed to evaluate the spectral resolution. Speckle patterns are captured at spectral intervals of 0.1 pm with an integration time of 300 $\mathrm{\upmu m}$. The ultra-stable laser used in the experiment is a fiber laser locked to a vacuum Fabry-Pérot (FP) cavity with a finesse of 400,000, employing the Pound-Drever-Hall frequency locking technique. The frequency drift of this laser is less than 100 kHz per day. The system uses active temperature control and passive vibration isolation, and the cavity is evacuated. Consequently, the lightwave frequency remains stable throughout the entire experiment, enabling ultra-high spectral resolution measurements. For long-term stability tests, the ultra-stable laser is also used as the light source to minimize frequency fluctuations and investigate the impact of environmental disturbances on speckle pattern distortion.

For broadband incident lightwave field reconstruction, broadband amplified spontaneous emission (ASE) noise generated by an erbium-doped fiber amplifier (EDFA, AEDFA-23-B-FA, Amonics) is used as the initial light source. The following waveshaper (WaveShaper 1000s, Finisar) is used to generate the required incident lightwave field. After passing through a 90:10 optical coupler, 10\% of the lightwave enters the OSA to obtain a reference spectrum for comparison, while the remaining 90\% enters the system for measurement. The reconstruction process is performed using a convex (CVX) optimization algorithm implemented in MATLAB. The personal computer used for computation is equipped with an AMD Ryzen 5 5600X CPU.

% \textcolor{red}{
% For optical frequency comb 
% 2 IQ Mach–Zehnder modulator (IQ-MZM, IQ-40-EVK, Realphoton Technology Co., Ltd )  Arbitrary Waveform Generator (AWG, M9502A, Keysight)
% }

\textbf{Reconstruction algorithm}

%写一下重构算法的原理的与实现过程。
% For arbitrary incident lightwave field, the resulting speckle pattern $I_n$ received by the charge-coupled device (CCD) can be mathematically expressed as:
% \begin{equation}
% \int_{\lambda_{\text {min }}}^{\lambda_{\text {max }}} T(\lambda) \cdot S(\lambda) d \lambda=I_n
% \label{eq3}
% \end{equation}
% where \textit{$T$} represents the spectral response of the proposed system. Due to practical limitations, the channel number is finite, and thus Equation (3) can be expressed as:
% \begin{equation}
% T_{N \times M} S_{M \times 1}=I_{N \times 1}
% \label{eq4}
% \end{equation}
% here \textit{$N$} denotes the number of spectral channels, and \textit{$M$} is the number of spatial channel.
% The inversion problem described above can be addressed using a nonlinear reconstruction algorithm, yielding a reconstruction result that closely matches the reference spectrum.

In computational spectropolarimetry, various reconstruction algorithms including traditional mathematical methods and emerging machine learning techniques are employed to recover the input lightwave field information from the acquired data. 
A commonly adopted practice is to perform a calibration process prior to measurements. For incident lightwave with known properties, the corresponding system response $T$, also named as transmission matrix is recorded, and a mapping relationship is established between the input and output, which serves as prior knowledge for the reconstruction. 
The aim of the reconstruction process is to produce results that closely match the reference information.
The reconstructed results $S$ can be obtained by solving a least squares linear regression problem:
\begin{equation}
S=\text { Minimize }\|I-TS\|_2
\label{eq5}
\end{equation}
where $I$ is the detected speckle.
Since noise is inevitably encountered in practical experiments, it becomes challenging to accurately reconstruct information using the standard least squares approach. 
To avoid overfitting problem and to improve reconstruction accuracy, a regularization-based reconstruction algorithm is employed.
The spectral and polarization information can be reconstructed by solving the nonlinear inverse problem described in Eq.(4):
\begin{equation}
S=\text { Minimize }\left(\|I-TS\|_2+\xi\|S\|_1+\zeta\|\Psi S\|_2\right)
\label{eq5}
\end{equation}
The first term $\|I-T S\|_2$ seeks the value of $S$ that best approximates the true spectrum under given constraints. 
The second term $\xi\|S\|_1$ serves as $L_1$-norms regularization to encourage sparsity in \textit{$S$}, while the third term $\zeta\|\Psi S\|_2$ acts as $L_2$-norm regularization to enforce smoothness in the reconstructed \textit{$S$}. Here, \textit{$\Psi$} is a discrete difference matrix.
The hyperparameters $\xi$ and $\zeta$ are the respective weighting coefficients for sparsity and broadband regularization, which are calculated using a standard $K$-fold cross-validation technique.
Through cross-validation, this method can be applied to different optical systems without requiring manual parameter tuning.
We adopt the convex (CVX) optimization framework to solve this optimization problem, which is grounded in compressive sensing theory. 
% Compared to other traditional matrix-based approaches, the CVX method offers higher reconstruction accuracy with relatively low computational cost, exhibits strong generalization performance, and is promising for enabling real-time measurement in future implementations. In fact not, it is just a common algorithm. We may should move this part to the Method below.

% \disclosures 

\bigskip

\textbf{Acknowledgements}
This work is financially supported by National Natural Science Foundation of China (NSFC) under Grant No. 62405178, 62435004.

\textbf{Data, Materials, and Code Availability} 
The data that support the findings of this study are available from the corresponding author upon reasonable request.

\textbf{Author contributions} 
Y. Wan and Q. Zhou conceived the idea, performed the experiments, analyzed the results and prepared the manuscript. 
L. Ma provided experimental equipment and assistance.
X. Fan and Z. He discussed the work and revised the paper.

\textbf{Conflict of interest} 
The authors declare no competing interests.

%%%%% References %%%%%

\bibliography{ref}   
\bibliographystyle{ieeetr}

%%%%% Biographies of authors %%%%%

% \vspace{2ex}\noindent\textbf{First Author} is an assistant professor at the University of Optical Engineering. He received his BS and MS degrees in physics from the University of Optics in 1985 and 1987, respectively, and his PhD degree in optics from the Institute of Technology in 1991.  He is the author of more than 50 journal papers and has written three book chapters. His current research interests include optical interconnects, holography, and optoelectronic systems. He is a member of SPIE.

% \vspace{1ex}
% \noindent Biographies and photographs of the other authors are not available.

% \listoffigures
% \listoftables

\end{spacing}
\end{document}